\documentclass[cits]{PoS}

\usepackage[authoryear,square]{natbib}
\bibpunct{(}{)}{;}{a}{}{,}

\title{SKA studies of {\it in-situ} synchrotron radiation from molecular clouds}

\ShortTitle{SKA studies of synchrotron radiation from molecular clouds}

\author{Clive Dickinson,$\!^{1}$\footnote{E-mail: Clive.Dickinson@manchester.ac.uk}~~~R.\,Beck,$\!^{2}$ R.\,Crocker$\,^{3}$ R.\,M.\,Crutcher,$\!^{4}$ R.\,D.\,Davies,$\!^{1}$ K.\,Ferriere,$\!^{5}$ G.\,Fuller,$\!^{1}$ T.\,Jaffe,$\!^{5}$ D.\,I.\,Jones,$\!^6$ J.\,P.\,Leahy,$\!^1$ E.\,J.\,Murphy,$\!^7$ M.\,W.\,Peel,$\!^1$ E.\,Orlando,$\!^8$ T.\,Porter,$\!^8$ R.\,J.\,Protheroe,$\!^9$ T.\,Robishaw,$\!^{10}$ A.\,W.\,Strong,$\!^{11}$ R.\,A.\,Watson,$\!^{1}$ F.\,Yusef-Zadeh$^{12}$ \\ 
$^1$Jodrell Bank Centre for Astrophysics, The University of Manchester, UK\\
$^2$MPIfR, Bonn, Germany \\
$^3$Australian National University, Canberra, Australia \\
$^4$Department of Astronomy, University of Illinois, USA \\
$^5$IRAP, Universit\'{e} de Toulouse, CNRS, Toulouse, France \\
$^6$Department of Astrophysics, Radboud University, Nijmegen, The Netherlands \\ 
$^7$IPAC, Caltech, MC 220-6, Pasadena CA, 91125, USA\\
$^8$Kavli Institute, Stanford University, Stanford, CA 94305, USA \\
$^{9}$Dept of Physics, University of Adelaide, South Australia 5005, Australia \\
$^{10}$DRAO, Penticton, BC, V2A 6J9, Canada \\
$^{11}$MPE, Garching, Germany \\
$^{12}$Department of Physics and Astronomy/CIERA, Northwestern University, Evanston, IL., USA \\
}

\abstract{
Observations of the properties of dense molecular clouds are critical in understanding the process of star-formation. One of the most important, but least understood, is the role of the magnetic fields. We discuss the possibility of using high-resolution, high-sensitivity radio observations with the SKA to measure for the first time the {\it in-situ} synchrotron radiation from these molecular clouds. If the cosmic-ray (CR) particles penetrate clouds as expected, then we can measure the B-field strength directly using radio data. So far, this signature has never been detected from the collapsing clouds themselves and would be a unique probe of the magnetic field. Dense cores are typically $\sim0.05$\,pc in size, corresponding to $\sim$arcsec at $\sim$kpc distances, and flux density estimates are $\sim$\,mJy at 1\,GHz. The SKA should be able to readily detect directly, for the first time, along lines-of-sight that are not contaminated by thermal emission or complex foreground/background synchrotron emission. Polarised synchrotron may also be detectable providing additional information about the regular/turbulent fields.
}

\FullConference{
Advancing Astrophysics with the Square Kilometre Array\\
June 8-13, 2014\\
Giardini Naxos, Sicily, Italy}

\begin{document}

\section{Introduction}

Star formation is one of the key processes in the Universe and is of fundamental importance to astrophysics \citep{McKee2007}. We still do not have a full understanding of star formation and, in particular, on the role that magnetic fields play \citep{Crutcher2012,Li2014}. Cosmic Rays (CRs) are the main ionising and heating agent in dense, starless, molecular cloud cores \citep{Padovani2013}. The role of magnetic fields has been debated for decades, but due to a lack of precise measurements it has still not been conclusively settled. There are two major classes of star-formation theories: i) strong field models where the B-field controls the molecular cloud, with ambipolar diffusion driving the formation and collapse of dense cores, and ii) weak-field models where turbulent flows trigger star formation. Although B-fields are at least comparable in strength to the turbulent pressure in the ISM, there is no definitive evidence for B-fields dominating gravity or for ambipolar-diffusion-driven star formation \citep{Crutcher2012}.  

It is very difficult to measure the detailed properties of dense molecular clouds, and perhaps the most difficult property is the magnetic field - its strength and geometry. The most common method is via Zeeman splitting of radio-frequency HI, OH, and CN lines or masers, which typically gives only the magnetic field strength along the line-of-sight, $B_{\rm los}$ \citep{Crutcher2012}\footnote{It is possible to get the total B-field strength when complete line splitting is observed, which is possible with masers \citep{Crutcher1999}; see also \cite{Robishaw2014}.}. Other methods are also difficult, which include using measurements of optical and near-infrared polarisation (extinction along the line-of-sight) and sub-mm polarised thermal dust emission (difficult from the ground) \citep{Poidevin2013}.

An alternative method of probing the magnetic field is via synchrotron radiation \citep{Brown1977,Marscher1978,Orlando2013}, which has rarely been mentioned in the literature in relation to molecular clouds. Synchrotron radiation is produced primarily by relativistic CR electrons when decelerated by magnetic fields. The intensity of synchrotron radiation depends only on the number and energy spectrum of CR electrons and the magnetic field strength perpendicular to the line-of-sight. We know the measured energy spectrum of CRs \citep{Ackermann2012}, at least on average at energies of relevance to radio synchrotron emission ($\sim$GeV), and we know that the CR density varies slowly throughout the Galaxy and can penetrate dense molecular clouds at these energies and above \citep{Brown1977,Marscher1978,Umebayashi1981}.$\!$\footnote{There have been claims that GeV CRs cannot fully penetrate the densest clouds, thus suppressing the CR diffusion coefficient \citep{Jones2008,Protheroe2008,Jones2011}.} In molecular clouds, the magnetic field strength is much larger than in the ambient interstellar medium \citep{Crutcher1999}. Therefore, in principle, the synchrotron intensity should give a detectable signature, which could be used as a probe of the magnetic field. This is a direct way of measuring the total magnetic field strength, including the irregular (turbulent) component, which most other indicators (Zeeman, optical polarisation) are not directly sensitive to, since they measure the regular (Zeeman) or ordered (optical polarisation) field.\footnote{Other indirect measures of the random component exist, such as the Chandrasekhar-Fermi method, which uses the dispersion of the measured polarisation angles to probe the magnetic field in the plane of the sky \citep{Chandrasekhar1953,Watson2001,Crutcher2004}. See also \cite{Hildebrand2009} and references therein for further extensions.} Polarised synchrotron emission could provide additional information, including the ordered versus anisotropic random component and projected angle of the magnetic field on the sky.

It is therefore somewhat surprising that very little attention has been given to using synchrotron as a probe of molecular clouds. \cite{Jones2008} observed two nearby (3--4\,kpc) dense cold starless cores (G333.125--0.562 and IRAS15596--5301) with the Australian Telescope Compact Array (ATCA) at 1384 and 2368\, MHz, to try to detect secondary leptons.$\!$\footnote{In this article we focus on primary CR electrons, although the conversion into secondary leptons could be significant in some clouds \citep{Dogel1990,Protheroe2008}. Observations with the SKA could also provide useful constraints on secondaries.} They found upper limits of $\sim 0.5$\,mJy/beam and constrained the B-field strength to $B<500\,\mu$G. However, this is still compatible with the scaling of $|B|$ and $n_{\rm H}$ - more sensitivity is required. \cite{Protheroe2008,Jones2011} only found upper limits from Sgr\,B2 after subtraction of the dominant thermal emission. The only possible candidate so far is from the G0.13--0.13 molecular cloud detection, which was detected at 74\,MHz, with an associated CO hotspot \citep{Yusef-Zadeh2013}. However, a displacement between the radio position and molecular core suggests it could be from a different region of space.

These non-detections can be partly understood due to the relatively weak (typically mJy or less) signal that is expected to come from {\it in-situ} synchrotron emission inside the cloud itself. This is due to the fact that on large scales ($\sim 1$--10\,pc), the magnetic fields in clouds appear to be relatively weak ($\sim 10\,\mu$G) while strong fields are on scales much smaller than this ($\sim 0.05$\,pc) resulting in a weak flux signal. Also, most of the collapsing clouds ("cores") are located at low latitudes where there is significant confusion from background synchrotron and free-free emission. Nevertheless, high resolution and high sensitivity observations could allow molecular clouds to be mapped in some sight-lines. This may also shed light on CR penetration into the densest clouds, which sometimes appear as a radio dark cloud (RDC) \citep{Yusef-Zadeh2012}. Note that recent high-resolution 5/20\,GHz JVLA observations of the Galactic centre cloud G0.216+0.016 have detected compact ($<2.2$\,arcsec, or sub-pc) non-thermal sources, which may be the signature of in-situ synchrotron radiation from secondary CR electrons \citep{Jones2014}.

In this chapter we briefly review the physics of synchrotron radiation and magnetic fields, and the relation that appears to exist between them in molecular clouds. We then give the prospects of detecting this signature in molecular clouds with the SKA.


\section{Synchrotron radiation}
\label{sec:synch}

Synchrotron radiation is emitted primarily by relativistic cosmic-ray electrons spiralling in the Galactic magnetic field. It is this radiation that often dominates the radio sky at frequencies below a few GHz. The theory of synchrotron radiation is well understood. For a power-law distribution of electron energies,

\begin{equation}
N(E) dE = N_0 E^{-\gamma} dE ~,
\end{equation}
the emissivity, $j_{\nu}$, of synchrotron radiation is given by

\begin{equation}
j_{\nu} \propto N_0 B^{(\gamma+1)/2} \nu^{(1-\gamma)/2} ~, 
\end{equation}
where $B$ is the magnetic field strength, $N_0$ is the number density of electrons, $\gamma$ is the power-law index of electron energies, and $\nu$ is the observing frequency. Highly relativistic (GeV and above) CR electrons are expected to penetrate dense clouds freely \citep{Umebayashi1981}. If the electron energy spectrum is a power-law with slope $-\gamma$ then the observed synchrotron radio emission spectrum is also a power-law with slope $\alpha=(1-\gamma)/2$ (flux density $S \propto \nu^{\alpha}$). The cosmic ray energy distribution at energies of order GeV can be approximated by a power-law with slope $\gamma \approx +2.5$--3.0 \citep{Ackermann2012}, which corresponds to a synchrotron index $\alpha \approx -0.8$. This is indeed the typical spectral index observed at GHz frequencies \citep{Reich1988,Platania1998}.

It can also be seen that the emissivity scales as $B^{(\gamma+1)/2}$, which means it goes as approximately $B^2$. This is of relevance to molecular cloud collapse, since the magnetic field is expected to be significantly amplified during collapse, and thus could give a detectable signal from {\it in-situ} synchrotron radiation.


\section{Magnetic fields in collapsing clouds}
\label{sec:bfields}

The role that magnetic fields play in the processes of molecular cloud and star formation has been debated for decades. Theoretical studies suggest that magnetic fields play an important if not crucial role in the evolution of interstellar clouds and the formation of stars. In summary, magnetic fields provide magnetic support against cloud collapse. There are various models of star formation and the details of the magnetic field are always important. For example, the core of a cloud can become unstable due to ambipolar diffusion, collapsing to form stars, while the envelope can remain in place. The connection between the core and the surrounding envelope by magnetic field lines can transfer angular momentum outward and make it possible for stars to form. Other star formation models have the dissipation of magnetised turbulence as a controlling factor in star formation. Measuring the magnetic field is a key observation that allows us to infer i) whether supersonic motions are Alfvenic, and ii) the relative importance of the gravitational, kinetic and magnetic densities in dense clouds \citep{Crutcher1999}. These observables thus allow us to test star formation models such as ambipolar diffusion and turbulence \citep{Crutcher2012,Lazarian2012}.

\begin{figure}[!h]
\centering
\includegraphics[width=.6\textwidth]{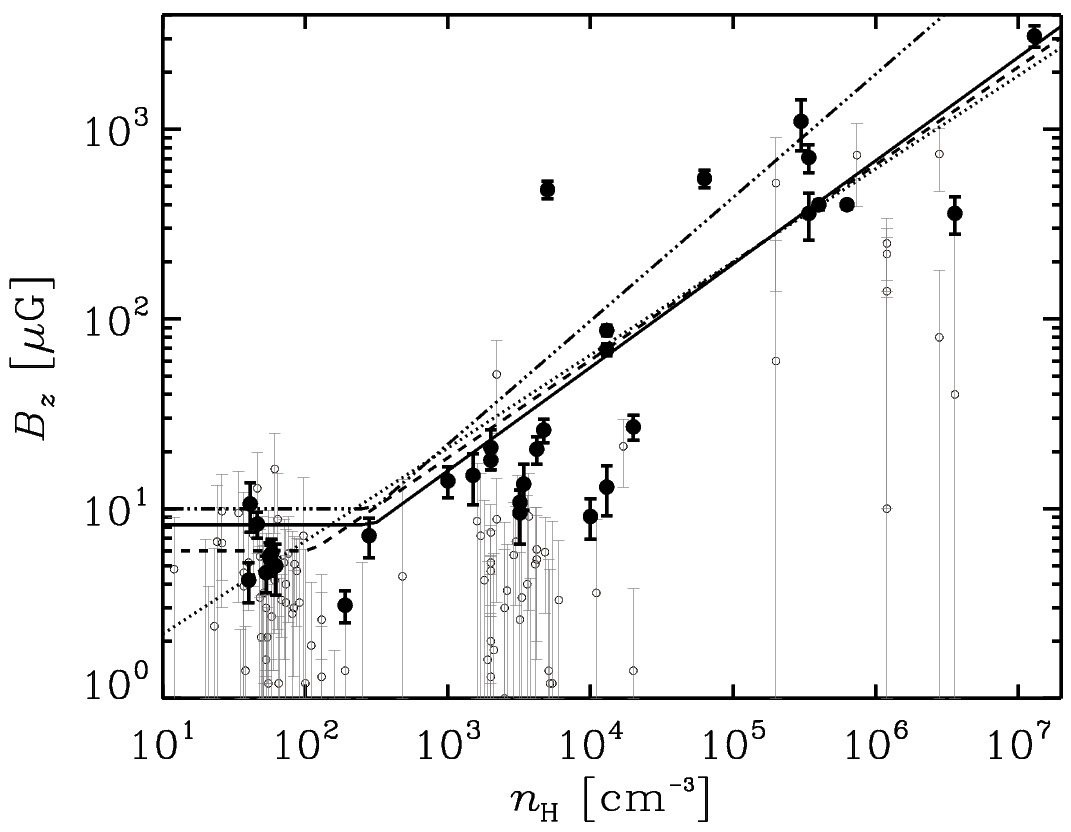}
\caption{Zeeman splitting measurements of the magnetic field of a number of molecular clouds, plotted against the volume density of molecular gas. Significant ($>3\sigma$) detections are shown as black filled circles (data taken from \citealt{Crutcher2010}). There is considerable scatter, yielding various slopes (see overplotted lines) depending on the exact model being fitted. The power-law slope between $B_z$ and density $n_{\rm H}$ is in the range $\kappa \approx 0.5$--0.7. Our best fit (solid line) for detections greater than $3\sigma$ significance yields $\kappa=0.54 \pm 0.05$ above $n_{\rm H}=300$\,cm$^{-3}$. Preliminary figure reproduced from a forthcoming publication \citep{Strong2015}.}
\label{fig:zeeman}
\end{figure}

Detailed measurements of the magnetic field strength and alignment are difficult. However, in recent years, direct measurements of the magnetic field strength have been made. Most notable are Zeeman splitting data \citep{Crutcher1999,Crutcher2010} and also sub-mm thermal dust emission \citep{Poidevin2013}. Detailed studies of Zeeman splitting from a sample of molecular clouds indicate that the thermal-to-magnetic pressure $\beta_p \approx 0.04$, implying that magnetic fields are important.  Moreover, the measurements showed that magnetic field strengths scale with gas densities as $B \propto n^{\kappa} \approx 0.5$---0.7, as shown in Fig.~\ref{fig:zeeman}. This is close to the theoretical value $\kappa=0.47$ predicted by models of ambipolar diffusion \citep{Fiedler1993}. The latest value appears to be $\kappa=0.65$ \citep{Crutcher2012} but there is considerable scatter in the measurement (Fig.~\ref{fig:zeeman}); our best-fitting value applied to detections above $3\sigma$ is $\kappa=0.54\pm0.05$, although there could be biases when neglecting non-detections \citep{Crutcher2010}. The large scatter may be related to the fact that Zeeman splitting is only sensitive to the regular (ordered and directional) magnetic field component along the line-of-sight; the $B$--$n_{\rm H}$ relation may be different for turbulent fields. Furthermore, this trend only occurs above some density $n_0 \sim 300$\,cm$^{-3}$, although this has still to be determined precisely. Clearly more data, and complementary probes of the magnetic field, are needed to make progress in this area.

\section{Predictions for synchrotron radiation from collapsing clouds}
\label{sec:lines}

Given that the synchrotron emissivity scales as $\sim B^2$ and $B$ scales as $\sim n_{\rm H}^{0.6}$, it is logical that it should also scale roughly as the volume density i.e. $j_{\nu} \propto n_{\rm H}$. From this, one might expect low frequency maps such as the Haslam et al. 408\,MHz map \citep{Haslam1982} to be bright around giant molecular clouds (GMCs) and for molecular clouds to be very bright in high resolution observations (e.g. VLA, ATCA). We will now use the observed scaling relation of B with $n_{\rm H}$ to estimate the flux density expected for typical molecular clouds. 

We assume that the ambient CR electrons pervade molecular clouds unimpeded and a power-law distribution of CR electron energies with slope $\gamma$ and a power-law slope between density $n_{\rm H}$ and B-field strength $B$ above a value $n_0=300$\,cm$^{-3}$. Using the CR flux model of \cite{Strong2011}, the predicted brightness temperature (in mK) at 408\,MHz can be be approximated by \citep{Strong2015}:

\begin{equation}
\left(\frac{T_{\rm pred}}{{\rm mK}}\right) = 2.8\times10^3  \left( \frac{N_{\rm H}}{10^{23}\,{\rm cm^{-2}}}\right) \left( \frac{n_{\rm H}}{300\,{\rm cm^{-3}}} \right)^{\kappa(\gamma+1)/2-1} ~,
\end{equation}

\noindent where $N_{\rm H}$ is the column density (cm$^{-2}$) and $n_{\rm H}$ the volume density (cm$^{-3}$). This corresponds to a predicted integrated flux density (in mJy) at 1\,GHz, for a source subtending a solid angle $\Omega_{\rm src}$, 

\begin{equation}
\label{eq:b_n}
\left(\frac{S_{\rm pred}}{{\rm mJy}}\right) = 6.6 \times 10^{6}  ~\left(\frac{\Omega_{\rm src}}{\rm sr}\right) \left( \frac{N_{\rm H}}{10^{23}\,{\rm cm^{-2}}}\right) \left( \frac{n_{\rm H}}{300\,{\rm cm^{-3}}} \right)^{\kappa(\gamma+1)/2-1}~.
\end{equation}

Table~\ref{tab:clouds} lists some example molecular clouds, using data from \cite{Crutcher1999}, with predicted flux densities at 1\,GHz. We have used the $B-n_{\rm H}$ relation above with $\kappa=0.6$, assume a synchrotron frequency spectral index $\alpha=1.0$ ($\gamma=+3.0$), and $\Omega_{\rm src}=\pi/4 \times \theta^2$. Dense molecular clouds have typical densities of $10^5$--$10^6$\,cm$^{-3}$ in H$_{2}$ and linear sizes of $\sim 0.05$\,pc. This gives column densities of $\sim 10^{23}$\,cm$^{-2}$. For typical distances of a $\sim$kpc, this corresponds to angular sizes of $\sim 10$\,arcsec. It can be seen that many of these sources have predicted flux densities of $\sim$mJy. It is interesting to see that a few sources have much larger predicted flux densities (e.g. Sgr\,B2 at about 1\,Jy). However, one has to be careful since these are due to the large physical size assumed (22\,pc for Sgr\,B2). In practice, the collapsing clouds tend to be very small, often clustered, in a parent cloud that is much larger. The magnetic field measured in the densest regions is unlikely to apply to the entire cloud. Thus it is easy to over-estimate the flux density in this way and this appears to be why dense molecular clouds are not bright in low resolution radio surveys such as the \cite{Haslam1982} 408\,MHz map. On the other hand, additional synchrotron from secondary leptons could boost the synchrotron level \citep{Protheroe2008}.

Therefore, the predicted flux densities should only be considered order-of-magnitude estimates at this point since the precise values depend very sensitively on the choice of $\kappa$ and $n_0$ and on the observed input parameters. Furthermore, the huge scatter about this relation observed in Fig.~\ref{fig:zeeman} already indicates that either the measurements are not representative of the mean field, or, the simple $B-n_{\rm H}$ relationship does not hold. New observations will be crucial for testing this hypothesis.

\begin{table*}
\centering
\caption{Molecular cloud data from \protect\cite{Crutcher1999} with estimates of predicted integrated synchrotron flux density $S_{\rm GHz}$ (mJy) at 1\,GHz based on the statistical $B$--$n_{\rm H}$ scaling law (equation~\protect\ref{eq:b_n}) with $\kappa=0.6$. The flux density has been scaled to 1\,GHz assuming a spectral index $\alpha=-0.8$ ($\gamma=2.6$). The brightness temperature $T_{\rm GHz}$ is what would be observed with a 51\,arcmin beam.}
\begin{tabular}{lccccccc}
\hline
Name  &$B_{z}$   &$n_{{\rm H}_2}$   &$R$    &D  &$\theta$ &$T_{\rm GHz}$ &$S_{\rm GHz}$   \\
    &[$\mu G$]  &[cm$^{-3}$]  &[pc]   &[kpc]  &[arcsec]    &[mK]   &[mJy]    \\ \hline
W3\,OH	   	&  3100 	&$6.31\times10^6$ &      0.02 	&       2.0 &       4.0 	&     0.06 		&      0.05 \\
DR21\,OH1     	&   710 	&$2.00\times10^6$ &      0.05 	&       1.8 &      11.2 	&      0.30 		&      0.27	 \\ 
Sgr B2	   	&   480 	&$2.51\times10^3$ &     22.0 	&       7.9 &    1149 	&    1200	 	&     1000 \\
M17\,SW	   	&   450 	&$3.16\times10^4$ &      1.0 		&       1.8 &     236 	&      31	 	&       27.0 \\
W3 (main)      	&   400 	&$3.16\times10^5$ &      0.12 	&       2.0 &      24.3 	&      0.49 		&      0.43 \\
S106	   		&   400 	&$2.00\times10^5$ &      0.07 	&       0.6 &      48.1 	&      0.74 		&      0.65 \\
DR21\,OH2      	&   360 	&$1.00\times10^6$ &      0.05 	&       1.8 &      11.2 	&      0.14 		&      0.13 \\
OMC-1	   	&   360 	&$7.94\times10^5$ &      0.05 	&       0.4 &      50.3 	&       2.3 		&        2.0 \\
NGC2024        	&    87 	&$1.00\times10^5$ &      0.2 		&       0.4 &     196 	&      14.6 		&       13.0 \\
W40                	&    14 	&$5.01\times10^2$ &      0.05 	&       0.6 &      34.4 	&     0.04 		&      0.03 \\
$\rho$\,Oph\,1 	&    10 	&$1.58\times10^4$ &      0.03 	&       0.1 &      91.7 	&      0.14 		&      0.13 \\
\hline
\end{tabular}
\label{tab:clouds}
\end{table*}


\section{Prospects for the SKA}

The high sensitivity and high resolution of the SKA should finally provide the first definitive detection of {\it in-situ} synchrotron radiation from molecular clouds themselves. The dense cores inside clouds are typically $0.05$\,pc in size, corresponding to $\sim 1$\,arcsec at a distance of  $\sim$kpc. This is well matched to the core of SKA1-MID, which provides good $u,v$ coverage (and therefore good surface brightness sensitivity) up to $\sim 5$\,arcsec. Estimated flux densities are of order mJy for typical dense molecular clouds at GHz frequencies (see Table~\ref{tab:clouds}), which is approximately the depth to which current observations have been targeted (e.g. \citealt{Jones2008}).  Optimal frequencies are in the range of a few hundred MHz (band 1--2 of SKA1-MID) to $\sim 2$\,GHz (band 3 of SKA1-MID). The very lowest frequencies ($\lesssim 300$\,MHz) will be affected by synchrotron and free-free self-absorption. Above a few GHz, the steep spectrum ($\alpha \approx -0.8$) means the signal will be very weak and more susceptible to free-free contamination, which has a flatter spectral index ($\alpha \approx -0.1$ in the optically thin regime). The SKA is therefore the ideal instrument for studying this, yet untapped, area of radio astronomy. 

Multi-frequency (matched $u,v$ coverage) continuum mapping will be important to verify that the emission is non-thermal. In principle, optically thin thermal emission could be removed using radio recombination lines, which could be detected using the wide-band spectral capability of the SKA. Mapping the polarisation of the synchrotron radiation provides further information on the magnetic field, such as the orientation and orderliness. The polarisation fraction could be up to $\approx 75\,\%$ for a ordered field with no depolarisation. However, in practice the observed level will be lower, making this a more difficult measurement to make. Lower frequencies (bands 1/2) will also likely be affected by Faraday depolarisation along the line-of-sight.

Diffuse clouds (e.g., \citealt{Heiles1997,Myers1995}) will often not be bound structures, have modest magnetic fields and an in-situ synchrotron signal will not be obvious due to confusion from nearby objects. Given their large size (arcmin to several degrees) and diffuse nature, they will not provide the cleanest signal and are not well-suited to SKA baselines. More well-defined clouds such as CO/sub-mm catalogues of molecular clouds (e.g., \citealt{Roman-Duval2010,Planck2011_XXIII}) provide lists of hundreds of clouds but, due to the relatively low angular resolution, they are relatively large in size (5--20\,pc) and may not contain the densest clouds of interest here. Proto-stellar outflows may also emit appreciable synchrotron radiation \citep{Sokoloski2008} which will need to be carefully mapped. Nevertheless, with the large field-of-view of the SKA, it should be possible to find useful areas (Galactic plane and molecular clouds, particularly at high Galactic latitude where cleaner sight-lines should exist) to survey to look for the signature of {\it in-situ} synchrotron emission.

The best targets will be the most compact and dense clouds, where the magnetic fields are likely to be highest, and the signal will be the cleanest. Examples such as those in Table~\ref{tab:clouds}, with angular sizes of $\sim$arcsec are ideal for the SKA. An example would be the ring of dense molecular cores in the W40 complex, where upper limits of $\sim$mJy exist but require deeper observations \citep{Pirogov2013}. The long baselines of SKA will in fact allow us to resolve the emission inside a dense cloud opening up the potential of studying the magnetic field structure within the cloud. Surveying a number of clouds in this way is likely to directly shed light on models of star formation through the alignment of the magnetic field. They should ideally have as little or no free-free emission from ionised gas or nearby supernova remnants (SNRs).\footnote{There is a class of SNRs that interact with molecular clouds and they generally show OH masers at the site of the interaction. These sources could be interesting targets for future observations since CRs interact with the cold gas and should produce synchrotron emission \citep{Wardle2002}.} In many cases, such surveys could piggy-back on other SKA surveys of the Galaxy.

The high density and quiescent nature of infrared dark clouds (e.g., \citealt{Peretto2009}), together with high resolution column density images available, will make them interesting targets for SKA observations to probe the magnetic field at early stages of formation. A good example would be the massive star-forming cores within the infrared dark cloud SDC335.579-0.272 \citep{Peretto2013}.


\section{Conclusions and outlook}
\label{sec:conclusions}

Observations of the properties of dense molecular clouds are critical in understanding the process of star-formation. One of the most important, but least understood, is the role of the magnetic fields. We propose to use high resolution, high sensitivity radio observations with the SKA to measure the {\it in-situ} synchrotron radiation from these molecular clouds, complementing other methods such as Zeeman splitting \citep{Robishaw2014}. If the CR particles penetrate as expected, then we can measure the B-field strength directly using radio data (the CR flux is relatively smooth and varies slowly throughout the Galaxy with a scale-height of $\approx 1$\,kpc; \citealt{Orlando2013,Stepanov2014}). If they cannot, then the ability of the SKA to pick up extended, low surface-brightness emission may allow us to search for synchrotron emission from the outer regions of such clouds.
Collapsing cores are typically arcsec in size and flux density estimates are $\sim$\,mJy at 1\,GHz and thus should be readily detectable by the SKA. The large field-of-view will allow many lines-of-sight to be investigated for a given molecular cloud. Multiple frequencies and radio recombination lines will allow separation from free-free emission while polarised data opens up the possibility of mapping the field geometry within collapsing clouds.

CD acknowledges support from an ERC Starting (Consolidator) Grant (no.~307209).

%





\bibliography{clive_refs}{}
\bibliographystyle{apj}


\end{document}